\documentclass[lettersize,journal]{IEEEtran}
\usepackage{amsmath,amsfonts}
\usepackage{algorithmic}
\usepackage{array}
\usepackage[caption=false,font=normalsize,labelfont=sf,textfont=sf]{subfig}
\usepackage{textcomp}
\usepackage{stfloats}
\usepackage{url}
\usepackage{verbatim}
\usepackage{graphicx}
\usepackage{multirow}
\usepackage{booktabs}
\usepackage{tabularx}
\usepackage{bbding}
\usepackage{cite}
\usepackage{hyperref}
\usepackage{url}

\def\BibTeX{{\rm B\kern-.05em{\sc i\kern-.025em b}\kern-.08em
    T\kern-.1667em\lower.7ex\hbox{E}\kern-.125emX}}
\usepackage{balance}

\begin{document}
\title{Multiscale Motion-Aware and Spatial-Temporal-Channel Contextual Coding Network for Learned Video Compression}

\author{Yiming Wang, ~\IEEEmembership{Student Member,~IEEE,} 
Qian Huang, ~\IEEEmembership{Member,~IEEE,} 
Bin Tang, ~\IEEEmembership{Member,~IEEE,} \\ 
Huashan Sun, and Xing Li

\thanks{Manuscript created October, 2023;  ({Corresponding author: Qian Huang})}

\thanks{Yiming Wang, Qian Huang, Bin Tang, Huashan Sun are with the school of computer and information, Hohai university, Nanjing, Jiangsu, 211100, China (email: isymwang@gmail.com, huangqian@hhu.edu.cn, cstb@hhu.edu.cn, sunhuashan@hhu.edu.cn).
}
\thanks{Xing Li is with College of information Science and Technology, Nanjing Forestry University, Nanjing 210037, China (email: lixing@njfu.edu.cn) }
}

\markboth{Journal of \LaTeX\ Class Files,~Vol.~18, No.~9, August~2023}%
{Motion-Aware and Spatial-Temporal-Channel Coding Network for Learned Video Compression}

\maketitle

\begin{abstract}
Recently, learned video compression has achieved exciting performance. Following the traditional hybrid prediction coding framework, most learned methods generally adopt the motion estimation motion compensation (MEMC) method to remove inter-frame redundancy. However, inaccurate  motion vector (MV) usually lead to the distortion of reconstructed frame. In addition, most  approaches ignore the  spatial and channel redundancy. To solve above  problems, we propose a motion-aware and spatial-temporal-channel contextual coding based video compression network (MASTC-VC), which learns the latent representation and uses  variational autoencoders (VAEs) to capture the characteristics of intra-frame pixels and inter-frame motion. Specifically, we design a multiscale motion-aware module (MS-MAM)  to estimate spatial-temporal-channel consistent motion vector by utilizing the multiscale motion prediction information in a coarse-to-fine way. On the top of it, we further propose a spatial-temporal-channel contextual module (STCCM), which explores the correlation of  latent representation to reduce the bit consumption from spatial, temporal and channel aspects respectively. Comprehensive experiments show that our proposed  MASTC-VC is surprior to previous state-of-the-art (SOTA) methods on three  public benchmark datasets. More specifically, our method brings  average 10.15\% BD-rate savings against H.265/HEVC (HM-16.20) in PSNR  metric and  average 23.93\% BD-rate savings against  H.266/VVC (VTM-13.2) in MS-SSIM metric. 
\end{abstract}

\begin{IEEEkeywords}
Learned video compression, motion estimation motion compensation,  multiscale  motion-aware module, spatial-temporal-channel contextual module, MASTC-VC.
\end{IEEEkeywords}

\section{Introduction}
\IEEEPARstart{W}{ith}  the rise of video applications such as YouTube and TikTok, short videos have become an extremely important media representation. People gradually have high demands on video resolution (eg., 1080p, 2K, 4K and even 8K videos) and smoothness, which are all dependent on video compression.  Over the past twenty years, several recognized  video compression standards including H.264/AVC \cite{DBLP:journals/tcsv/WiegandSBL03}, H.265/HEVC \cite{DBLP:journals/tcsv/SullivanOHW12} and H.266/VVC \cite{DBLP:journals/tcsv/BrossWYLCSO21}  have been developed by the Joint Video Group (JVT). However, higher coding performance is often followed  by an exponential increase of coding complexity and these hand-crafted modules cannot be jointly optimized in an end-to-end fashion.

In recent years, deep learning techniques are booming and developing rapidly.  Ballé \emph{et al.} \cite{DBLP:journals/corr/BalleLS16a} first introduce variational autoencoders to build an end-to-end learned image compression framework.  Based on \cite{DBLP:journals/corr/BalleLS16a},  many subsequent works \cite{DBLP:conf/iclr/BalleMSHJ18, DBLP:conf/nips/MinnenBT18, DBLP:conf/iclr/LeeCB19, DBLP:conf/cvpr/ChengSTK20, DBLP:conf/cvpr/HeZSWQ21,DBLP:conf/cvpr/HeYPMQW22} are proposed to make significant progress in improving rate distortion performance, even outperforming the best current hand-crafted video coding standard  H.266/VVC \cite{DBLP:journals/tcsv/BrossWYLCSO21} for intra-frame (I-frame) coding. 
The success of learned  image compression has driven the rapid rise of learned  video compression. Similar to classical hybrid coding frameworks, the learned video compression approaches replace traditional modules  with deep neural networks for end-to-end rate distortion optimization (RDO). However, it is more challenging because both spatial and temporal redundancies need to be reduced.

From the coding perspective, current learned video compression methods 
 \cite{DBLP:conf/cvpr/LuO0ZCG19, DBLP:conf/cvpr/Lin0L020, DBLP:conf/eccv/LuCZCOXG20, DBLP:conf/cvpr/AgustssonMJBHT20,DBLP:conf/eccv/HuCXLOG20, DBLP:journals/pami/LuZO0G021, DBLP:conf/cvpr/HuL021, DBLP:conf/nips/LiLL21, DBLP:conf/mm/GaoC00ZZ22, DBLP:conf/eccv/HoCCGP22, DBLP:journals/tcsv/LinJZWMG23, DBLP:journals/tip/JinLPPLL23,Yang,DBLP:journals/tcsv/LiuLMWXCW21, DBLP:journals/tcsv/LiuLCCMW22} are roughly classified into two  categories.  \textbf{Motion prediction based method}: motion prediction utilizes the correlations between  frames to get predicted information in pixel domain or feature domain.  \textbf{Entropy coding based method}:  advanced entropy model aims to capture spatial-temporal correlations to reduce redundancy. The above approaches correspond to two main aspects of optimization: 1) how to estimate accurate motion vector (MV)  for prediction. 2) how to design efficient entropy coding models.  

To estimate  accurate motion vector, many  methods have been proposed, such as  pixel-level optical flow based methods \cite{DBLP:conf/cvpr/LuO0ZCG19,  DBLP:conf/eccv/LuCZCOXG20,DBLP:conf/eccv/HuCXLOG20, DBLP:journals/pami/LuZO0G021},   space-space flow based method \cite{DBLP:conf/cvpr/AgustssonMJBHT20},  multiple-frame method \cite{DBLP:conf/cvpr/Lin0L020}, feature-level prediction methods \cite{DBLP:conf/cvpr/HuL021,DBLP:conf/mm/GaoC00ZZ22} and convolutional LSTM (ConvLSTM) \cite{NIPS2015_07563a3f} based method \cite{DBLP:journals/tcsv/LinJZWMG23}.  However,  optical flow based methods \cite{DBLP:conf/cvpr/LuO0ZCG19,  DBLP:conf/eccv/LuCZCOXG20,DBLP:conf/eccv/HuCXLOG20, DBLP:journals/pami/LuZO0G021} rely on dense flow for explicit learning,  which makes it difficult to extract accurate motion information and may introduce extra artifacts in  complicated non-rigid scenario.  Faced with this situation, space-space flow based method \cite{DBLP:conf/cvpr/AgustssonMJBHT20} is proposed to extend the 2D flow field to the 3D space, leading to excessive spatial-temporal correlations. Multi-frame method \cite{DBLP:conf/cvpr/Lin0L020} is also  proposed to reduce this redundancy, but at the expense of complex motion prediction network. Feature-level prediction methods \cite{DBLP:conf/cvpr/HuL021,DBLP:conf/mm/GaoC00ZZ22} employ implicit learning, which mainly uses unsupervised learning to reconstruct the MV, avoiding the limitations introduced by optical flow. Nevertheless, it only uses the single-scale motion estimation strategy and ignores the spatial structure of motion field, resulting in less efficient motion representation. ConvLSTM based method \cite{DBLP:journals/tcsv/LinJZWMG23} takes spatial-temporal sequence prediction to model inter-frame motion and obtains progressive performance, however brings the burden of  encoding side.  

Meanwhile, several entropy coding models are designed to reduce redundancy. It can be divided into the residual coding based scheme \cite{DBLP:conf/cvpr/LuO0ZCG19, DBLP:conf/cvpr/Lin0L020, DBLP:conf/eccv/LuCZCOXG20, DBLP:conf/cvpr/AgustssonMJBHT20, DBLP:journals/pami/LuZO0G021, DBLP:conf/cvpr/HuL021, DBLP:journals/tcsv/LiuLCCMW22, DBLP:journals/tcsv/LinJZWMG23} and conditional coding based scheme \cite{DBLP:conf/nips/LiLL21, DBLP:conf/eccv/HoCCGP22, DBLP:conf/mm/GaoC00ZZ22, DBLP:journals/tip/JinLPPLL23,DBLP:journals/tcsv/LiuLMWXCW21}.  Among them, residual coding based scheme is suboptimal and uses only hand-crafted subtraction operations to remove redundancy.  In contrast,  conditional coding based scheme uses temporal context for entropy modeling and introduce hyper priors to reduce redundancy by transforming  the marginal distribution model of coded symbols into a optimized joint model.  However, current contextual models  ignore the spatial context and channel context.  This is because the hyper priors and contexts can be combined to predict the entropy model, which approximates the latent probability distribution more accurately.
 
In our work,  we propose a  motion-aware and spatial-temporal-channel contextual coding based video compression network (MASTC-VC). More specifically, we introduce a multiscale motion-aware module (MS-MAM) that introduces the coarse-to-fine strategy  and utilizes the multiscale motion prediction information to  extract spatial-temporal-channel consistent motion representations for more accurate MV.  Furthermore, based on the contextual coding architecture, we design a spatial-temporal-channel  contextual module (STCCM) to learn more precise spatial context, temporal context and channel context. Finally, we aggregate three contexts to enhance the prediction capabilities of the entropy model.

\textbf{Our contributions can be summarized as follows}:
\begin{itemize}
\item We propose a multiscale motion-aware module (MS-MAM), which explores the multiscale motion prediction information and performs coarse-to-fine strategy to estimate spatial-temporal-channel consistent MV. 

\item We propose a spatial-temporal-channel contextual module (STCCM) that aggregates spatial context, temporal context and channel context to  reduce the bit-rate effectively.

\item Extensive experimental results demonstrate that our proposed MASTC-VC outperforms the state-of-the-art (SOTA) methods on HEVC, UVG and MCL-JCV benchmark datasets. 

\end{itemize}

\section{Related Work}
\subsection{Learned image compression}
Compared with traditional image compression standards \cite{125072,DBLP:journals/spm/SkodrasCE01, bpg}, learned image compression methods \cite{DBLP:journals/corr/BalleLS16a, DBLP:conf/iclr/BalleMSHJ18, DBLP:conf/nips/MinnenBT18, DBLP:conf/iclr/LeeCB19,DBLP:conf/cvpr/ChengSTK20, DBLP:conf/cvpr/HeZSWQ21,DBLP:conf/cvpr/HeYPMQW22} have received more attention and  reconstructed high-quality images while maintaining low bit rates. Mainly because these methods typically employ the auto-encoder framework that transforms input images into latent space to generate compact representations and design advanced entropy models to estimate their probability distributions.

In \cite{DBLP:journals/corr/BalleLS16a},  Ballé \emph{et al.}  first propose a learned image compression network,  which solves the backpropagation of quantization operations and takes a factorized entropy model to approximate the latent space distribution. Then,  Ballé \emph{et al.} \cite{DBLP:conf/iclr/BalleMSHJ18} further introduce hyper priors that are seen as the side-information to model spatial correlation, achieving results comparable to HEVC intra-frame coding \cite{DBLP:journals/tcsv/SullivanOHW12}.  Subsequent Minnen \emph{et al.} \cite{DBLP:conf/nips/MinnenBT18}  and Lee \emph{et al.} \cite{DBLP:conf/iclr/LeeCB19} propose autoregressive models to construct a Gaussian mixture model (GMM), which outperforms the HEVC intra-frame coding.  However, autoregressive model is performed sequentially for all spatial locations and is therefore quite slow.  He \emph{et al.} \cite{DBLP:conf/cvpr/HeZSWQ21} propose two-step coding to solve the sequential coding problem without sacrificing performance. Cheng  \emph{et al.} \cite{DBLP:conf/cvpr/ChengSTK20} also propose a hybrid Gaussian model with results comparable to VVC intra-frame coding \cite{DBLP:journals/tcsv/SullivanOHW12}.

\subsection{Learned video compression}

The progress of learned image compression has also driven the growth  of learned video compression.

\textbf{Motion prediction based method.}  The pioneering DVC \cite{DBLP:conf/cvpr/LuO0ZCG19} adopts a existing optical flow estimation network to perform temporal prediction. Afterwards, there have been some improvements in motion prediction coding. For example, 
Lin \emph{et al.} \cite{DBLP:conf/cvpr/Lin0L020} extend single reference frame to multiple frames to obtain more accurate MV.  Agustsson  \emph{et al.} \cite{DBLP:conf/cvpr/AgustssonMJBHT20} propose a scale space flow estimation method, which 
use means of Gaussian blurring to transform 2D optical flow to 3D spaces. Hu \emph{et al.} \cite{DBLP:conf/eccv/HuCXLOG20} propose resolution-adaptive flow coding method, where the optimal motion resolution is determined  by the RDO strategy.  Liu \emph{et al.}  \cite{DBLP:journals/tcsv/LiuLMWXCW21} use motion representations of multi-scale flow field with joint spatio-temporal prior aggregation for motion coding. However, the optical flow estimation network is designed to generate accurate motion maps, which may not be optimal for video compression tasks. In addition, the above flow-based methods typically employ bilinear warping, which may introduces additional artifacts.  
 
Subsequent FVC \cite{DBLP:conf/cvpr/HuL021} is proposed to use deformable convolution (DCN) \cite{DCN} to predict motion patterns in the feature space. Furthermore, Gao \emph{et al.} \cite{DBLP:conf/mm/GaoC00ZZ22} utilize the previous raw  frame as an auxiliary reference  information  for motion prediction.   Lin \emph{et al.} \cite{DBLP:journals/tcsv/LinJZWMG23} 
 propose a ConvLSTM based method that  explores more reference frames to model the motion representations.   In general, accurate MV is crucial  for motion prediction. For learned video compression methods, it is preferable for  MV to be spatial-temporal-channel consistent.

\textbf{Entropy coding based method.} Another research direction is entropy coding.  residual coding based schemes  \cite{DBLP:conf/cvpr/LuO0ZCG19, DBLP:conf/cvpr/Lin0L020, DBLP:conf/eccv/LuCZCOXG20, DBLP:conf/cvpr/AgustssonMJBHT20, DBLP:journals/pami/LuZO0G021, DBLP:conf/cvpr/HuL021, DBLP:journals/tcsv/LiuLCCMW22, DBLP:journals/tcsv/LinJZWMG23} use simple subtraction operations to remove redundancy between successive frames.  Recently, conditional coding based scheme gradually attracts more and more attention, due to the fact that conditional coding consumes fewer bits.  Li \emph{et al.} \cite{DBLP:conf/nips/LiLL21} firstly move from residual coding to conditional coding,  which learn temporal context and explore inter-frame correlation to  remove redundancy automatically. Yang  \emph{et al.} \cite{Yang} propose the recurrent probability entropy model which takes full advantage of temporal correlation to achieve efficient performance. Sheng  \emph{et al.} \cite{TCM} design the feature propagation to learn temporal context more efficiently. The following work \cite{DBLP:conf/eccv/HoCCGP22} further introduces conditional augmented normalizing flow to form a purified conditional coding structure. Jin  \emph{et al.} \cite{DBLP:journals/tip/JinLPPLL23} also utilize multi-frequency components in temporal context  for entropy coding.  

However, the above conditional coding variants mainly utilize the temporal context and drop the spatial context, mainly because the autoregressive model is non-parallel for spatial correlation. Moreover, all methods also ignore information  in the channel dimension.

\begin{figure*}[htbp]
\centering
\includegraphics[width=\linewidth]{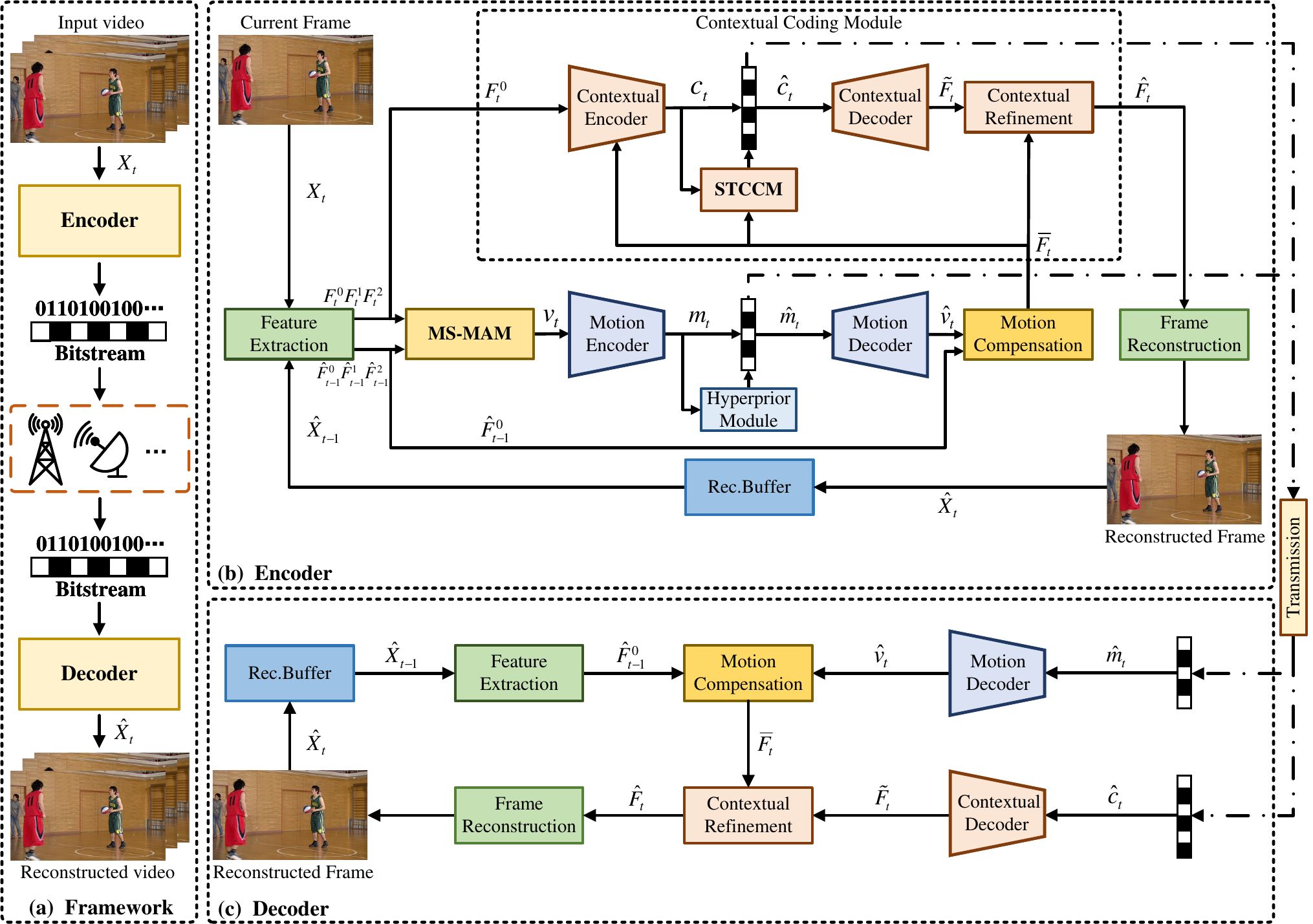}%
\vspace{-2mm}
\caption{Overview of our proposed video compression framework (MASTC-VC).  (a) Framework: the encoder encodes input video to get bitstream which is transmitted over the internet, the decoder receives the bitstream and decodes it to get compressed video. (b) Encoder: it mainly uses the multiscale motion-aware module (MS-MAM) to estimate the motion vector and the spatial-temporal-channel contextual module (STCCM) to reduce the entropy bit consumption for generating the bitstream. (c) Decoder: it receives the bitstream and decodes it to get the final reconstructed frame.}
\vspace{-3mm}
\end{figure*}

\begin{figure}[htbp]
\centering
\includegraphics[width=0.8\linewidth]{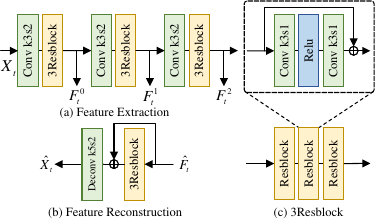}%
\vspace{-2mm}
\caption{Network structure of (a) our feature extraction module and (b) our feature reconstruction module with the details of resblocks are shown in (c).}
\vspace{-3mm}
\end{figure}

\section{METHODOLOGY}
\subsection{Framework Overview}
  Given raw video sequence $X=\left\{X_1, X_2, \ldots, X_{t-1}, X_t\right\}$,  we intend  to generate high-quality reconstructed frames $\hat X=\left\{\hat{X_1}, \hat{X_2}, \ldots, \hat X_{t-1}, \hat{X_t}\right\}$ with lower bit consumption, where  $X_t$ and $\hat{X_t}$  denote the current  and reconstructed frame at the time step $t$, respectively.  Hence, we propose a  motion prediction compensation framework (MASTC-VC) mainly including a multi-scale motion-aware module (MS-MAM) and spatial-temporal-channel contextual module (STCCM).  The workflow of our proposed method is described in Fig.1.
 
 \textbf{1) Feature Extraction}: Our feature extraction module is shown as Fig.2(a), which transforms the pixel-level current frame  $X_t$ and previous reconstructed frame $\hat X_{t-1}$ into the extracted multi-scale  feature $F_{t}^i$ and $\hat F_{t-1}^i$,  as follows: 
 
\vspace{-1mm}
\begin{equation}
F_{t}^i,\hat F_{t-1}^i=Feature\_Extract\left(X_{t}, \hat X_{t-1}\right), i=0,1,2
\end{equation}

 \textbf{2) Multiscale Motion-Aware Module}:  The MS-MAM explores the spatial structure, temporal coherence and channel adaptability of motion information  on extracted multi-scale features to get spatial-temporal-channel consistent MV $v_t$ in a coarse-to-fine fashion (See III-B).

 \textbf{3) Motion Encoder-Decoder }: Motion Encoder-Decoder module is proposed to encode and decode MV $v_t$  to  reconstruct motion information  $\hat{v_t}$ in Fig.4. Details are given in III-C.
 
\textbf{4) Motion Compensation}: We use DCN to warp the reconstructed MV  $\hat{v_t}$  with the reference frame feature $\hat F_{t-1}^0$ to generate predicted feature $\bar {F_t}$ in the motion compensation.  (See III-C).

\textbf{5) Contextual Coding Module}: The current frame feature $F_t^0$ and predicted feature $\bar {F_t}$  are as conditions to extract spatial, temporal and channel contexts rather than simple subtraction operations to get contextual feature $\tilde F_t$. Then it is refined to get reconstructed feature $\hat{F_t} $.  We will introduce it in III-D.

 \textbf{6) Frame Reconstruction}: As shown in Fig.2(b), the feature reconstruction with three resdiual blocks and a deconv layers will transform the final reconstructed feature  $\hat{F_t} $ into the  pixel-level frame $\hat{X_t}$, which is stored in the reconstructed buffer (Rec.Buffer) for the following iterations.

\begin{figure*}[htbp]
\centering
\includegraphics[width=0.8\linewidth]{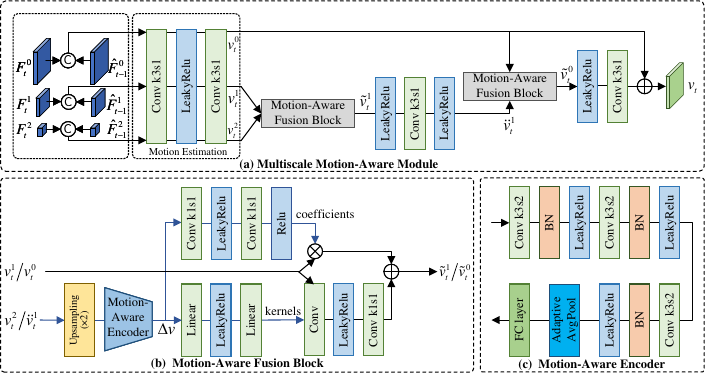}%
\vspace{-2mm}
\caption{The detailed network structures of (a) Multiscale Motion-Aware Module (MS-MAM), (b) Motion-Aware Fusion Block and (c) Motion-Aware Encoder.}
\vspace{-3mm}
\end{figure*}

\subsection{Multiscale Motion-Aware Module}
Considering that inaccurate  MV  will affect the subsequent motion prediction and entropy coding, we propose a multiscale motion-aware module (MS-MAM) as illustrated in Fig.3,  which performs motion estimation in a coarse-to-fine strategy.  Specifically,  we first concatenate multi-scale features $F_t^i$, and  $\hat F_{t-1}^i, i=0,1,2$ with the resolution of ($H\times W$, $\frac{1}{2} H \times \frac{1}{2} W$, $\frac{1}{4} H \times \frac{1}{4} W$) and fed  them into the motion estimation module to generate  initial temporal motion vectors ($v_t^i, i=0,1,2 $) with the size of ($H\times W$, $\frac{1}{2} H \times \frac{1}{2} W$, $\frac{1}{4} H \times \frac{1}{4} W$).

\vspace{-1mm}
\begin{equation}
\begin{aligned}
&v_t^i=Conv_{3\times 3} \cdot LeakyRelu\cdot Conv_{3\times 3} \cdot c(F_t^i, \hat F_{t-1}^i)\\
\end{aligned}
\end{equation}
where  $c$ denotes concatenation along channel dimension, $Conv_{3\times 3}$ represents the convolution operations with $3\times 3$  kernel  and $LeakyRelu$ denotes the activation function.

Furthermore, our motion-aware fusion block uses the motion representation as conditions to predict the kernel of the convolution at the  spatial level and  also generates modulation coefficients at the  channel level, respectively.  In detail, initial small scale MV $v_t^2$  is upsampled and then is encoded as motion-aware feature $\Delta v$ by motion-aware encoder.

\vspace{-1mm}
\begin{equation}
\Delta v= MAE (upsampling(v_t^2))
\end{equation}
 where MAE represents the motion-aware encoder operations.

The architecture of motion-aware encoder is presented  in Fig.3 (c) and it mainly consists of Convolution and BatchNorm layers. The final adaptive AvgPooling and Full Connected layers further integrate and utilize these feature representations to enable motion perception. On one hand, motion-aware feature $\Delta v$ is  fed into two linear layers to predict convolution kernels at the spatial level.

\vspace{-1mm}
\begin{equation}
kernels=Linear\cdot LeakyRelu\cdot Linear(\Delta v)
\end{equation}
where  kernels denotes 3 $\times $3  convolution kernels, Linear denotes a linear layer. 

On the other hand, motion-aware feature $\Delta v$ is fed into another two convolution  layer and two activation function to generate modulation coefficients to perform channel feature adaptation.

\vspace{-1mm}
\begin{equation}
coefficients =Relu \cdot Conv_{1 \times 1 }\cdot LeakyRelu \cdot Conv_{1 \times 1 }(\Delta v)
\end{equation}

where $Conv_{1 \times 1 }$ represents $1 \times 1$ convolution layer, $LeakRelu$ and $Relu$ denote the activation function.

Moreover, MV $v_t^1$ is rescaled from different channels by $coefficients$ to get channel components. MV $v_t^1$  is  processed with kernel-adaptive convolution  and 1 $\times$ 1 convolution layer to get spatial components. Finally, channel component features and  spatial component features are summed to obtain the coarse MV $\tilde v_t^1$ with $\frac{1}{2} H \times \frac{1}{2} W$ size.

\vspace{-1mm}
\begin{equation}
\begin{aligned}
\tilde v_t^1 & =     coefficients \bigotimes v_t^1 \\
             & + Conv_{1 \times 1 } \cdot LeakyRelu \cdot Conv_{kernels }(v_t^1)    
\end{aligned}            
\end{equation}
 where $\bigotimes$ denotes multiplication, $ Conv_{kernels }$ represents the convolution layer with the predicted kernels.

 Then, MV $\tilde v_t^1$ is passed to a series of operations including LeakyRelu activation function and convolution layer to generate the coarse MV$\ddot{v}_t^1$.
 
 \vspace{-1mm}
\begin{equation}
\ddot{v}_t^1=LeakyRelu \cdot Conv_{3 \times 3 } \cdot LeakyRelu(\tilde v_t^1 )
\end{equation}

Besides,  coarse MV $\ddot{v}_t^1$ with $\frac{1}{2} H \times \frac{1}{2} W$ size and $v_t^0$ with $ H \times  W$ resolution is fed into the motion-aware fusion block to generate  MV $\tilde v_t^0$ with $ H \times  W$ size. 

\vspace{-1mm}
\begin{equation}
\tilde v_t^0= Motion\mbox{-}Aware\_Fusion\_Block (\ddot{v}_t^1, v_t^0)
\end{equation}

Finally, $\tilde v_t^0$ is fed into a convolution layer and $v_t^0$ is summed to obtain a refined spatial-temporal-channel consistent MV $v_t$ with $ H \times  W$ size at the time step $t$.
\vspace{-1mm}
\begin{equation}
v_t = v_t^0+Conv_{3 \times 3 } \cdot LeakyRelu(\tilde v_t^0)
\end{equation}

\begin{figure}[t]
\centering
\includegraphics[width=0.6\linewidth]{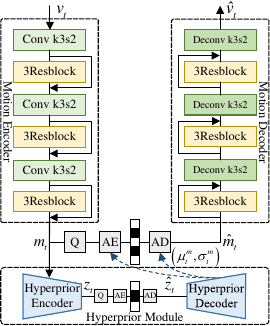}%
\vspace{-2mm}
\caption{Illustration of the motion encoder-decoder framework. Q represents quantization. AE and AD denote arithmetic encoder and arithmetic decoder. $\mu_t^m$ and $\sigma_t^m$ are the mean and variance value from hyperprior module.}
\vspace{-3mm}
\end{figure}

\subsection{Motion Encoder-Decoder and Motion Compensation}
After  MV $v_t$ is generated, it will be passed to  our proposed variational  auto-encoder style networks and then bitstream is transferred to the decoder side for motion compensation. The Motion Encoder-Decoder module is depicted in Fig.4. 
This module contains motion encoder and motion decoder that both are made up of a set of residual blocks \cite{7780459}. The architecture of residual blocks is shown in Fig.2(c).  MV $v_t$ is encoded as feature representation $m_t$ by motion encoder and $m_t$ is quantized as $\hat m_t$ which will be used for entropy coding. Furthermore, we use hyperprior module \cite{DBLP:conf/iclr/BalleMSHJ18} to estimate the mean $\mu_t^m$and variance  $\sigma_t^m$ value of Gaussian scale mixture.

\vspace{-1mm}
\begin{equation}
\left\{\begin{array}{l}
z_t =Hyperprior\_Encoder(m_t) \\
\mu_t^m,\sigma_t^m = Hyperprior\_Decoder(\hat z_t)\\
\end{array}\right.
\end{equation}

 where $z_t$ is the hyperprior feature representation and $\hat z_t$ denotes quantized  hyperprior feature representation.  
 
 We build entropy model $p_{\hat{m}}$ and fit the marginal probability distribution $q_{\hat{m}}$.  According to the Shannon entropy,  the encoding lower bound for modeling the feature representation using the estimated entropy model $p_{\hat{m}}$ is:
 
\vspace{-1mm}
\begin{equation}
\begin{aligned}
R\left( \hat m\right) &=\mathbb{E}_{\hat{m} \sim q_{\hat{m}}}\left[-\log _2 p_{\hat{m}}(\hat{m})\right] \\
&=-\sum_{t} q\left(\hat {m}_{t}\right) \log _2 {p}_{\hat{m}_{{t}}}\left(\hat m_{{t}}\right)
\end{aligned}
\end{equation}

where $q_{\hat{m}}$ denotes real probability distribution,  $p_{\hat{m}}$ is formulated as Gaussian scale mixture
parameterized by $(\mu_t^m,\sigma_t^m)$.  The above cross entropy can be more accurately optimized  when the probability distribution predicted by the entropy coding model $p_{\hat{m}}$  is the same as the actual probability distribution $q_{\hat{m}}$.  Total number of bits consumed by motion is the sum of $R(\hat m_t)$ and $R(\hat z_t)$ at the time step $t$.  Subsequently quantized features $\hat m_t$ is used  to reconstruct the MV $\hat v_t$ in the Motion Decoder.

The network architecture of motion compensation is shown in Fig.5. Specifically,  We utilize the reconstructed MV to compute the DCN offsets $ o_{t \rightarrow t-1}$  and modulation mask $m_{t \rightarrow t-1}$.

\begin{figure}[t]
\centering
\includegraphics[width=0.85\linewidth]{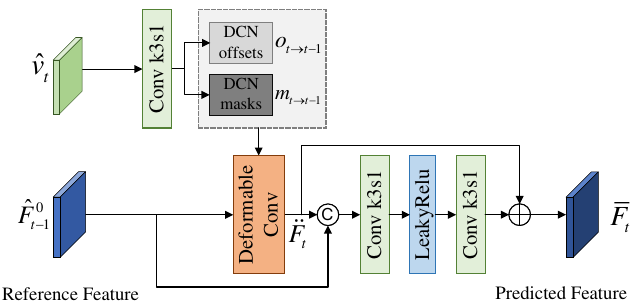}%
\vspace{-2mm}
\caption{Network architecture of Motion Compensation using deformable convolution. }
\vspace{-3mm}
\end{figure}

\vspace{-1mm}
\begin{equation}
o_{t \rightarrow t-1},  m_{t \rightarrow t-1}= Conv_{3 \times 3 }(\hat v_t)
\end{equation}
where  $o_{t \rightarrow t-1}$ represents DCN offsets, $m_{t \rightarrow t-1}$ represents DCN masks. DCN is then used to warp reconstructed MV $\hat v_t$  with the reference frame feature $\hat F_{t-1}^0$  with the resolution of $H\times W$ to produce  intermediate feature $\ddot {F_t}$. 

\vspace{-1mm}
\begin{equation}
\ddot{F}_t=DCN(o_{t \rightarrow t-1}, m_{t \rightarrow t-1}, \hat F_{t-1}^0)
\end{equation}

 Finally, we concatenate intermediate feature $\ddot{F}_t$ and reference feature $\hat F_{t-1}^0)$ to get predicted feature $\bar F_t$.
\vspace{-1mm}
\begin{equation}
\bar F_t=\ddot{F}_t+ Conv_{3 \times 3 } \cdot LeakyRelu \cdot Conv_{3 \times 3 }(c(\ddot{F}_t,  \hat F_{t-1}^0) )
\end{equation}

\begin{figure}[htbp]
\centering
\includegraphics[width=\linewidth]{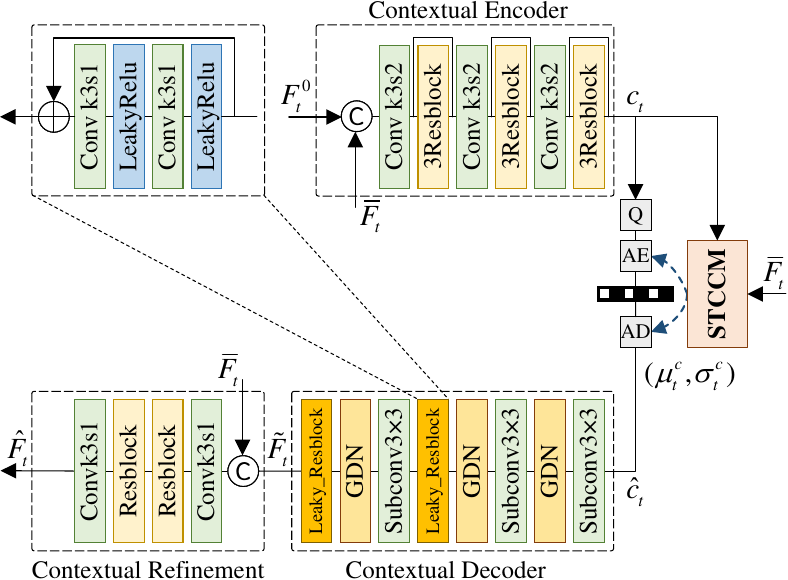}%
\vspace{-2mm}
\caption{Illustration of contextual coding module. STCCM is used to estimate  the probability distribution. }
\vspace{-3mm}
\end{figure}

\subsection{Contextual Coding Module}
Previous video compression methods mainly reduce redundancy  from temporal dimension, which have emerged with residual coding schemes and conditional coding based schemes. According to the information theory, the entropy of residual coding is equal to or larger than the contextual coding scheme. Therefore, we use conditional coding based scheme. Our contextual coding module mainly contains contextual encoder-decoder, refinement and STCCM, which is described in Fig.6.

The current feature $F_t^0$ and predicted feature $\bar F_t$ are concatenated to fed into the contexual encoder module to get contexual feature  representation $c_t$. 
\vspace{-1mm}
\begin{equation}
c_t=Contextual\_Encoder(c(F_t^0,\bar F_t))
\end{equation}

\begin{figure*}[t]
\centering
\includegraphics[width=0.8\linewidth]{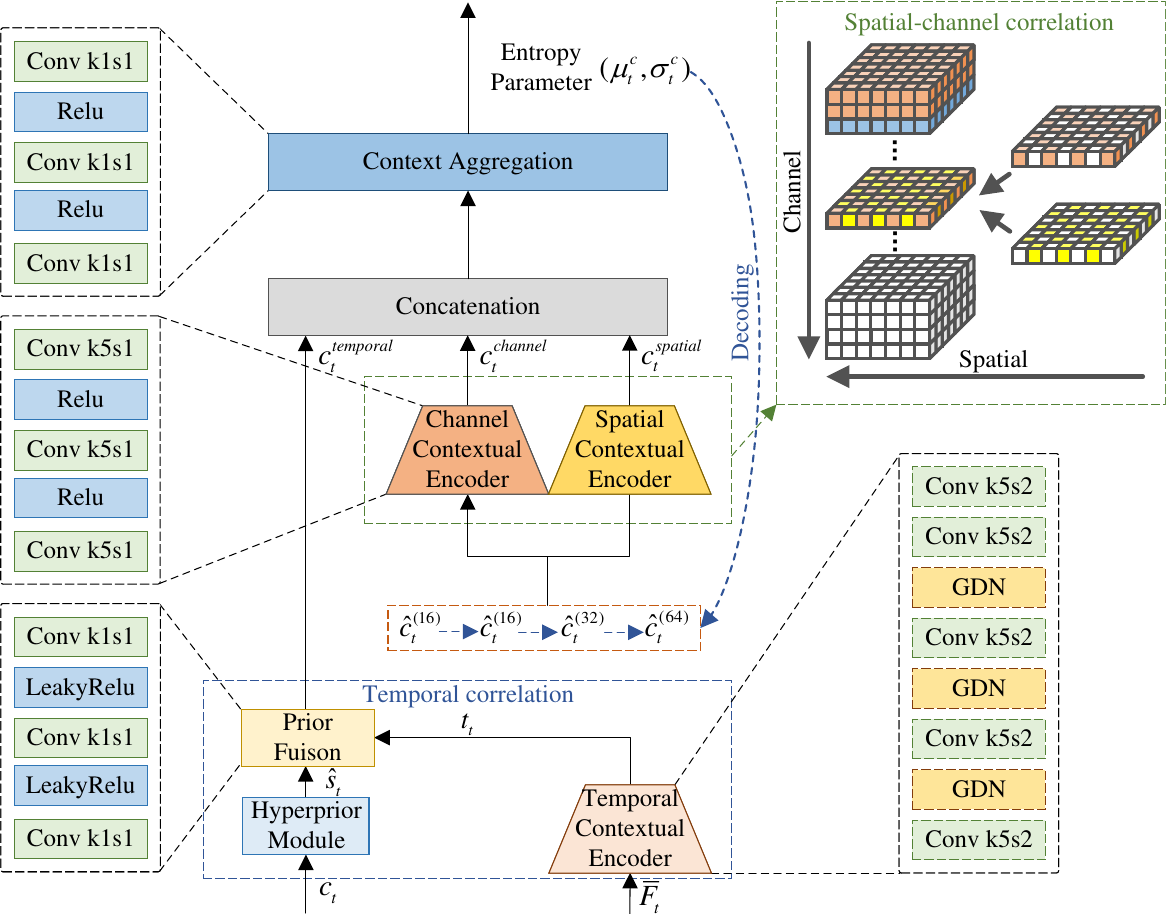}%
\vspace{-2mm}
\caption{Diagram of proposed spatial-temporal-channel contextual module (STCCM). Temporal, spatial and channel contexts are aggregated to estimate the entropy parameter.}
\vspace{-3mm}
\end{figure*}

\begin{figure}[htbp]
\centering
\includegraphics[width=0.8\linewidth]{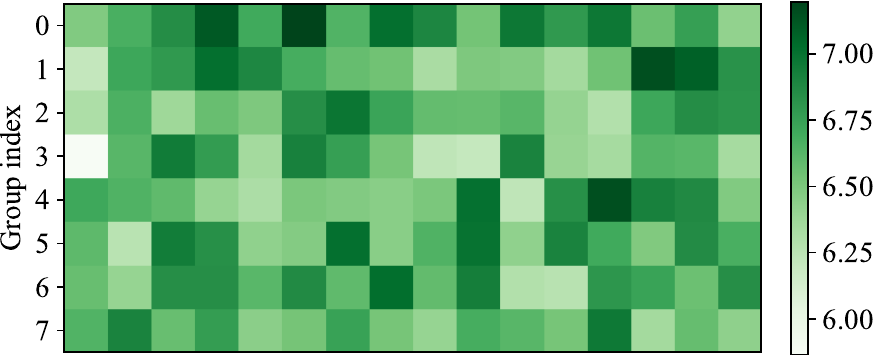}%
\vspace{-2mm}
\caption{Entropy of each feature channel. The results are evaluated on the HEVC Class B. Darker colors represent greater entropy.}
\end{figure}

We visualize the entropy of contexual feature  representation $c_t$.  From Fig.8, We find that there exists information aggregation in the contextual coding scheme for learned video compression and it is obvious to see that the entropy of previous four groups decreases gradually,  however this phenomenon does not apply to the last four groups.  This is because the previous four groups features are produced from the current uncompressed features encoded by contextual encoder, whereas the last four groups features are derived from the predicted features.  The current features are transformed from the pixel space to the latent space, just like the transformation of traditional video compression into the frequency domain, where the information is gathered in the initial channel. Besides, predicted features are generative and go through the motion encoder-decoder module, so we finally split the channel dimensions into 4 chunks with \{16, 16, 32, 64\} channels respectively. Subsequent ablation experiments will further verify the effectiveness of our proposed STCCM.

Fig.7 illustrates  our proposed spatial-temporal-channel contextual module (STCCM). We apply the hyperprior module  and temporal context model to extract hyperprior $\hat s_t$ and temporal prior $t_t$. Then  $\hat s_t$ and $t_t$ are fused to utilize temporal correlation and generate temporal context $c_t^{temporal}$. Furthermore, we introduce   spatial and  channel contextual encoder to explore spatial-channel correlation for the channel context $c_t^{channel}$ and spatial context $c_t^{spatial}$, respectively. To improve computational efficiency, we adopt the checkboard method \cite{DBLP:conf/cvpr/HeZSWQ21} for spatial correlation.

\vspace{-1mm}
\begin{equation}
\left\{\begin{array}{l}
\hat s_t =Hyperprior\_Module(c_t)\\
t_t =Temporal\_Contexutal\_Encoder(\bar F_t) \\
c_t^{temporal} = Prior\_Fusion(\hat s_t, t_t)\\
c_t^{channel} = Channel\_Contextual\_Encoder(\hat c_t^{(k)})\\
c_t^{channel} =Spatial\_Contextual\_Encoder(\hat c_t^{(k)}) \\
\end{array}\right.
\end{equation}

Three contexts information $c_t^{temporal}$, $c_t^{channel}$ and $c_t^{spatial}$ are concatenated and fed into context aggregation to predict entropy parameter ($\mu_t^c, \sigma_t^c$) of every chunk for the following decoding  $\hat c_t^{(k)}$.  Decoded $k$-dimensional feature channel  $\hat c_t^{(k)}$ will be used as conditions to extract the following channel context $c_t^{channel}$  and spatial context $c_t^{spatial}$ with temporal context $c_t^{temporal}$,  till reconstruct the entire contexual feature  representation $\hat c_t$.

Moreover, we build contextual entropy model $p_{\hat{c}}$ to fit probability distribution $q_{\hat{c}}$. According to the Shannon entropy,  the encoding lower bound for modelling the feature representation using the estimated entropy model $p_{\hat{c}}$ is:

\vspace{-1mm}
\begin{equation}
R\left( \hat c\right)=\mathbb{E}_{\hat{c} \sim q_{\hat{c}}}\left[-\log _2 p_{\hat{c}}(\hat{c})\right]=-\sum_{t} q\left(\hat {c}_{t}\right) \log _2 {p}_{\hat{c}_{{t}}}\left(\hat c_{{t}}\right)
\end{equation}
Therefore, the total number of bits consumed by context is the sum of $R(\hat c_t)$ and $R(\hat s_t)$ at the time step $t$.

Reconstructed context $\hat c_t$ is decoded as contextual feature $\tilde F_t$.  Then, $\tilde F_t$ and $\bar F_t$ are concatenated to be refined as reconstructed current feature $\hat F_t$. Finally, $\hat F_t$ is transformed into reconstructed frame $\hat X_t$.

\subsection{Loss Function}
Our proposed scheme aims to minimize the coding bitrate cost while reducing the distortion between current frame and reconstructed frame. We use the following rate-distortion (RD) loss function for training:

\vspace{-1mm}
\begin{equation}
\begin{aligned}
{L}_{t} & =\lambda D+R \\
& =\frac{1}{T} \sum_t\left\{\lambda d\left(X_t, \hat{X}_t\right)+R(\hat m_t)+R(\hat z_t)+R(\hat c_t)+R(\hat s_t)\right\}
\end{aligned}
\end{equation}

where $d(\cdot)$ denotes the mean-square-error (MSE) / 1- MS\mbox{-}SSIM \cite{1292216} for PSNR/MS\mbox{-}SSIM metric. $R(\hat m_t)$ denotes the encoding quantized motion vector feature representation and $R(\hat z_t)$ denotes the associated hyper prior of MV. $R(\hat c_t)$ denotes the encoding quantized contextual feature representation  and $R(\hat s_t)$ denotes the associated hyper prior of contextual feature. $\lambda$ denotes the Langrange  factor. $T$ is the time interval.

\section{EXPERIMENTS}
\subsection{Datasets and Experimental Setup}

\emph{\textbf{ Datasets}}: Following previous works, we employ  Vimeo-90K \cite{xue2019video} as a training dataset, which contains 89800 video clips and each clips have 7 successive frames with the resolution of 448$\times$256.  During the training phase, the videos are cropped with 256$\times$256$\times$3 size. Furthermore,  we use the HEVC datasets, UVG datasets \cite{DBLP:conf/mmsys/MercatVV20} and MCL\mbox{-}JCV \cite{MCL} datasets to measure video coding performance. Specifically, the HEVC datasets contains the video sequences with various resolutions of 1920$\times$1080 (Class B), 832$\times$480 (Class C), 416$\times$240 (Class D) and 1280$\times$720 (Class E). The Ultra Video Group (UVG) contains 7 high frame rate 1920$\times$1080 video sequences.  MCL-JCV dataset contains 30 1920$\times$1080 videos collected from YouTube. 

\emph{\textbf{Evaluation Methodologies}}: We adopt the bit per pixel (Bpp) to measure the bit consumption and use PSNR/MS\mbox{-}SSIM metrics to evaluate the distortion between the original frame and reconstructed frame. Besides, we  set the group of picture (GOP) size as 10 for HEVC datasets and 12 for other datasets as previous works settings. In addition, we set x265(veryslow) as the anchor used to report BD-rates.

\emph{\textbf{Implementation Details}}: We take a two-stage  strategy to train our model.  At the first stage,  we set $T$ as 1 in Eq.(18) and utilize successive frame information including I-frame and P-frame for 2,000K steps with the learning-rate of 5e-5 to achieve a baseline model. At the second stage, we take a widely-used training strategy \cite{content},  which sets $T$ as 6 in Eq.(18) and extends the length of training video sequences to 7 frames for another 500K steps with the learning-rate of 5e-6 to avoid error propagation.  When using MS\mbox{-}SSIM metric, we further fine-tune the model from stage 2 for 100K steps by using 1- MS\mbox{-}SSIM as distortion loss. Our proposed MASTC-VC is built by Pytorch 1.13 with CUDA 12.1. We set the batch size as 4 at the first stage and 2 at the remaining stage. We take 5 days to construct the entire experiments on the machine with a single NVIDIA 4090 GPU (24GB memory).

\begin{table*}[htbp]
\centering
\caption{BD-rate(\%) saving comparison between existing baseline methods and the anchor is x265(verslow) in psnr metrics. the best performance of learned methods is marked in bold.}
\label{tab:my-table}
\resizebox{\textwidth}{!}{%
\begin{tabular}{ccccccccccc}
\toprule[1.5pt]
\multirow{2}{*}{Datasets} & \multicolumn{2}{c}{Traditional Methods} & \multicolumn{7}{c}{Learned Methods}                                   \\ \cmidrule(lr{0pt}){2-3}\cmidrule(lr{0pt}){4-11}
 &
  \begin{tabular}[c]{@{}c@{}}HM-16.20 \cite{HM}\\ (LDP, 4refs)\end{tabular} &
  \begin{tabular}[c]{@{}c@{}}VTM-13.2 \cite{VTM}\\ (LDP, 8refs)\end{tabular} &
  \begin{tabular}[c]{@{}c@{}}FVC \cite{DBLP:conf/cvpr/HuL021}\\ (CVPR'21)\end{tabular} &
  \begin{tabular}[c]{@{}c@{}}DCVC \cite{DBLP:conf/nips/LiLL21}\\ (NIPS'21)\end{tabular} &
  \begin{tabular}[c]{@{}c@{}}SPME(FVC*) \cite{DBLP:conf/mm/GaoC00ZZ22}\\ (ACMMM'22)\end{tabular} &
  \begin{tabular}[c]{@{}c@{}}SPME(DCVC) \cite{DBLP:conf/mm/GaoC00ZZ22}\\ (ACMMM'22)\end{tabular} &
  \begin{tabular}[c]{@{}c@{}}CANF-VC \cite{DBLP:conf/eccv/HoCCGP22}\\ (ECCV'22)\end{tabular} &
  \begin{tabular}[c]{@{}c@{}}DMVC \cite{DBLP:journals/tcsv/LinJZWMG23}\\ (TCSVT'23)\end{tabular} &
  \begin{tabular}[c]{@{}c@{}}TCVC \cite{DBLP:journals/tip/JinLPPLL23}\\ (TIP'23)\end{tabular} &
  \textbf{Ours} \\ \midrule
HEVC ClassB               & -30.53                &    -53.23             & -15.83 & -32.04 & -23.25 & -35.56 & -33.15 & -32.12 &-40.52 & \textbf{-45.90} \\ \midrule
HEVC ClassC               & -18.59                &   -42.39              & -4.55  & -6.29  & -0.41  & -10.72 & -13.24 & -13.77 &-15.58 & \textbf{-15.87} \\ \midrule
HEVC ClassD               & -17.50                &   -40.34              & -1.51  & -11.74 & -6.22  & -15.48 & -15.68 & -18.51 &-24.08 & \textbf{-24.91} \\ \midrule
HEVC ClassE               & -46.91                &  -69.03               & -35.16 & -28.79 & -24.12 & -32.90 & -28.27 & -41.48 &-44.14 & \textbf{-48.05} \\ \midrule
UVG                       & -30.26                & -53.17                & -32.06 & -34.66 & -26.03 & -48.75 & -39.84 & -36.68 &-38.58 & \textbf{-54.54} \\ \midrule
MCL-JCV                   & -17.57                &   -40.44              & -17.44 & -20.32 & 1.25   & -30.30 & -22.17 & -14.81 &-21.07 & \textbf{-32.96} \\  \midrule
AVG                       & -26.89                &   -49.77              & -17.76 & -22.31 & -13.13 & -28.95 & -25.39 & -26.23 &-30.66 & \textbf{-37.04} \\  \bottomrule[1.5pt]
\end{tabular}%
}
\end{table*}

\begin{figure*}[htbp]
	\centering
	\begin{minipage}{0.32\linewidth}
		\centering
		\includegraphics[width=\linewidth]{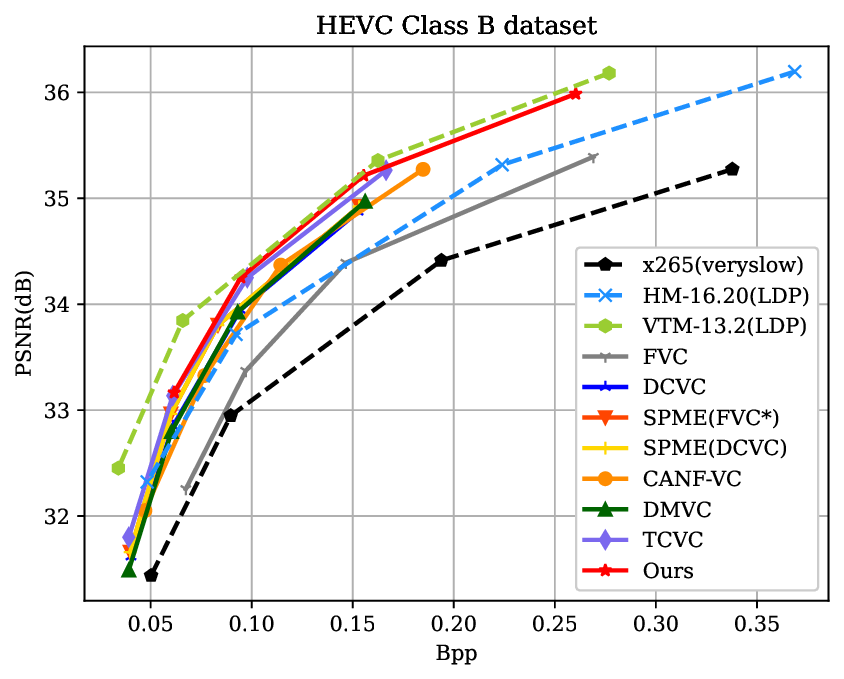}
		\label{chutian1}
	\end{minipage}
	\begin{minipage}{0.32\linewidth}
		\centering
		\includegraphics[width=\linewidth]{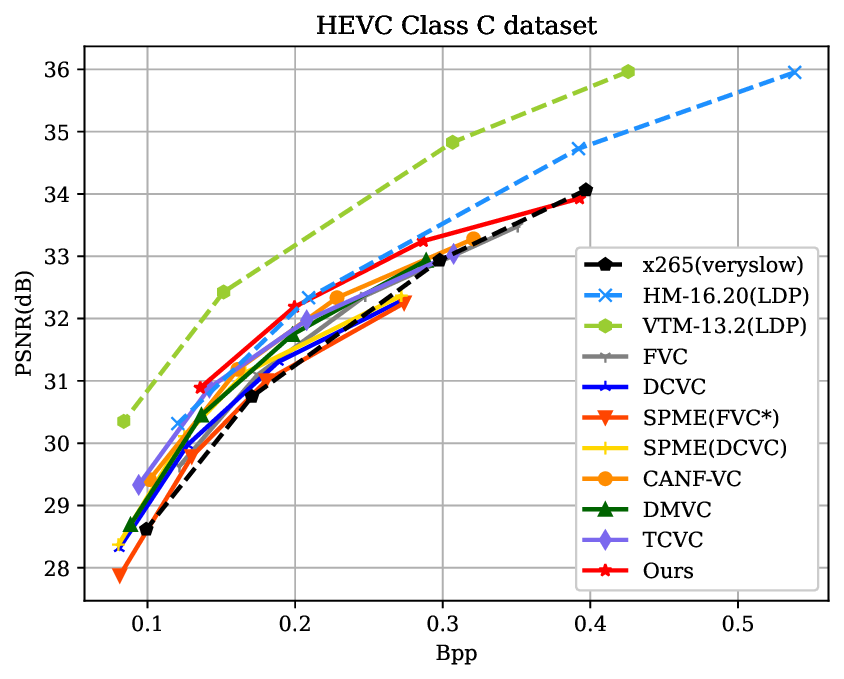}
		\label{chutian2}
	\end{minipage}
	\begin{minipage}{0.32\linewidth}
		\centering
		\includegraphics[width=\linewidth]{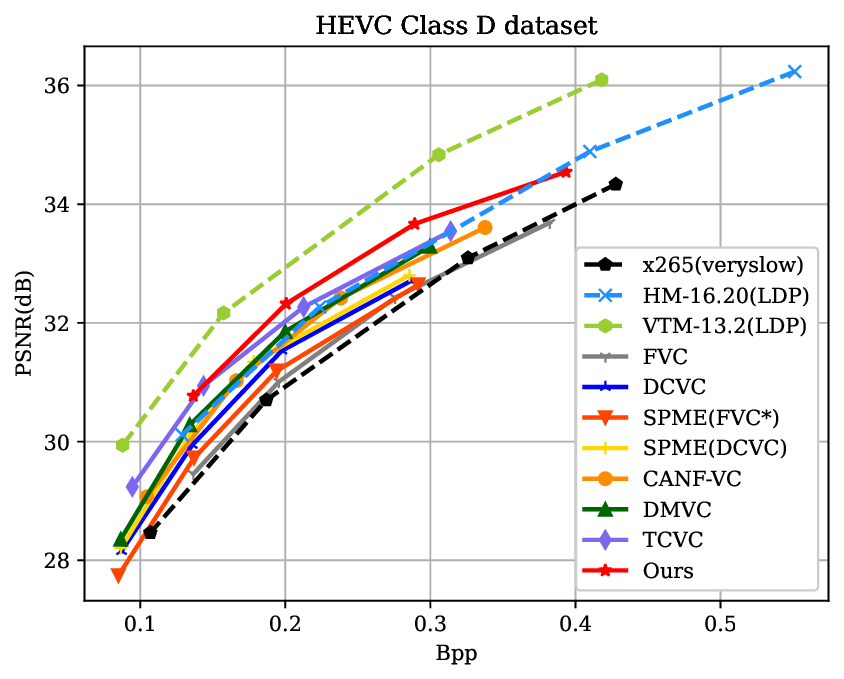}
		\label{chutian2}
	\end{minipage}
	\vspace{-2mm}
	\begin{minipage}{0.32\linewidth}
		\centering
		\includegraphics[width=\linewidth]{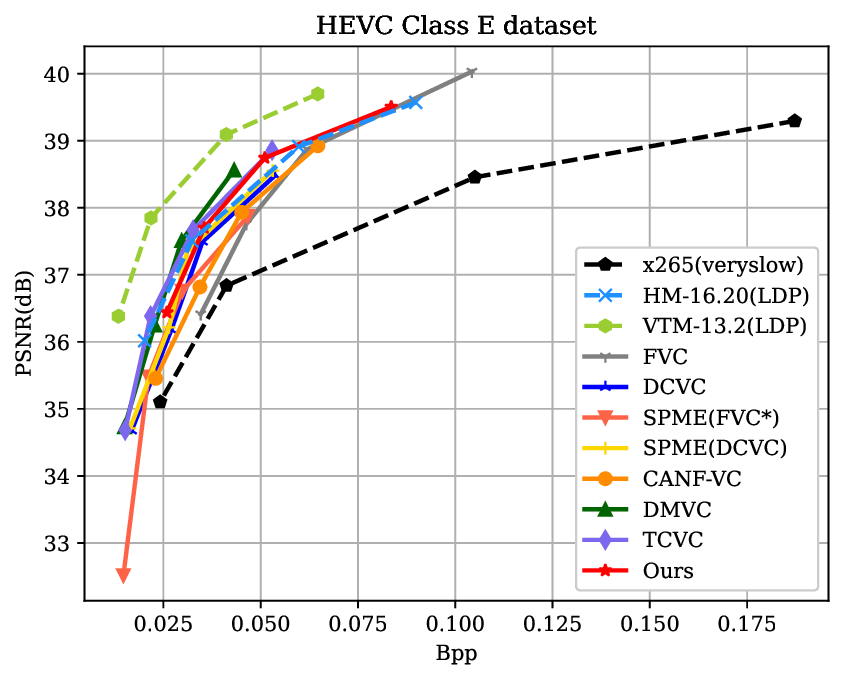}
		\label{chutian3}
	\end{minipage}
	\begin{minipage}{0.32\linewidth}
		\centering
		\includegraphics[width=\linewidth]{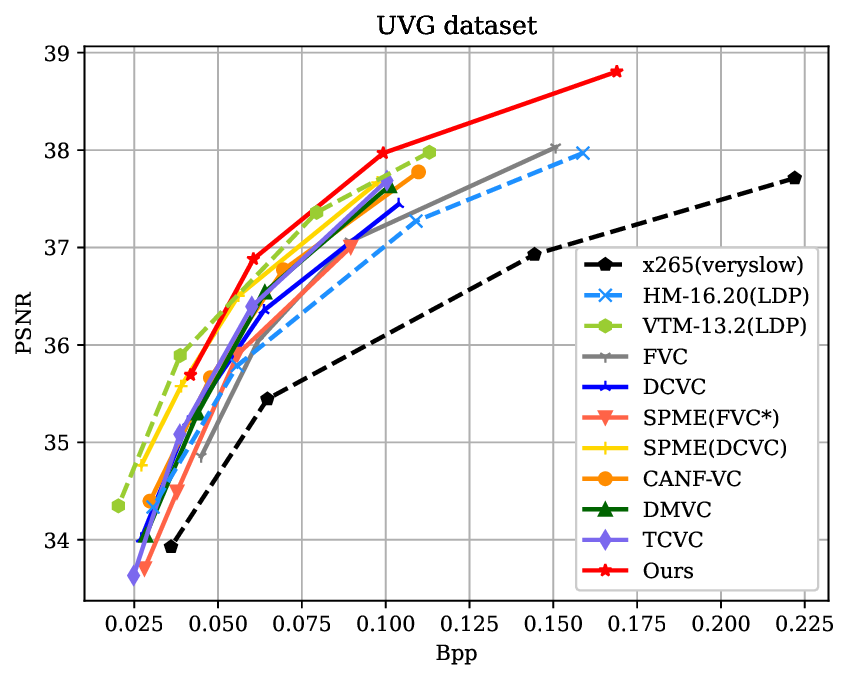}
		\label{chutian4}
	\end{minipage}
    \begin{minipage}{0.32\linewidth}
		\centering
		\includegraphics[width=\linewidth]{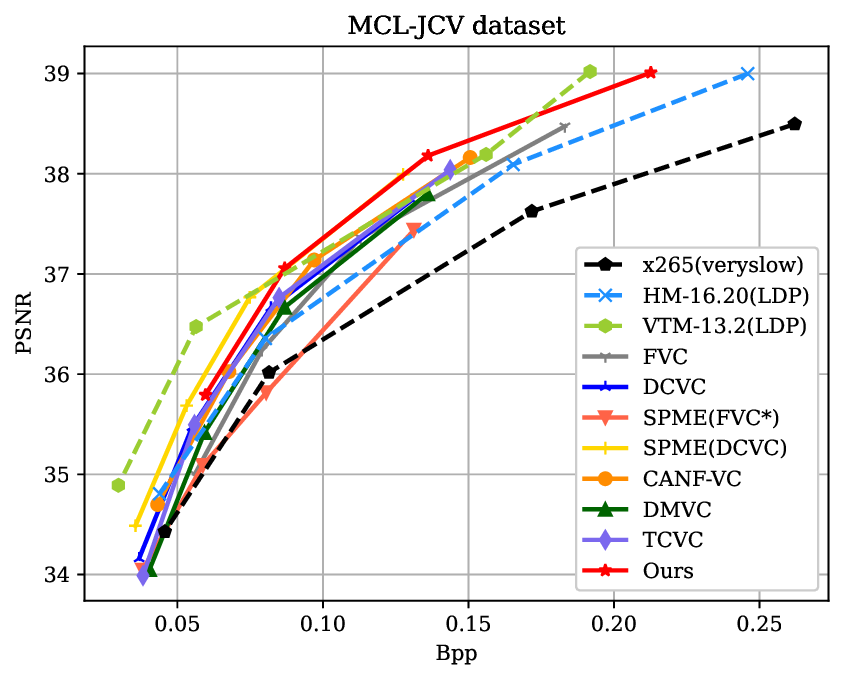}
		\label{chutian4}
	\end{minipage}

\caption{Rate-distortion performance evaluation of our proposed method on the HEVC, UVG and MCL-JCV datasets in PSNR metrics.}
\end{figure*}

\subsection{Experimental Results}
\emph{\textbf{Baseling Methods Settings:}} To evaluate the coding performance of our proposed MASTC-VC, we use traditional coding standards H.265/HEVC \cite{DBLP:journals/tcsv/SullivanOHW12}, H.266/VVC \cite{DBLP:journals/tcsv/BrossWYLCSO21} as well as the SOTA learned video compression methods including FVC \cite{DBLP:conf/cvpr/HuL021}, DCVC \cite{DBLP:conf/nips/LiLL21}, SPME(FVC*) \cite{DBLP:conf/mm/GaoC00ZZ22}, SPME(DCVC) \cite{DBLP:conf/mm/GaoC00ZZ22}, CANF-VC \cite{DBLP:conf/eccv/HoCCGP22}, DMVC \cite{DBLP:journals/tcsv/LinJZWMG23} and TCVC \cite{DBLP:journals/tip/JinLPPLL23}. Following previous work, we use the \emph{Cheng2020\_Anchor(MSE/MS-SSIM)} \cite{DBLP:conf/cvpr/ChengSTK20} provided by CompressAI \cite{begaint2020compressai} for I frame coding. The other prediction (P) frames are coded sequentially by their individual network.

For the current popular coding standsrd H.265/HEVC \cite{DBLP:journals/tcsv/SullivanOHW12}, we use the industrial software x265 (veryslow) and official reference software HM-16.20 (LDP) \cite{HM} as:

\begin{itemize}

\item x265 (verslow)

\texttt{ffmpeg -pixfmt yuv420p -s WxH -r FR 
-i Input.yuv -v frames N -c:v libx265 -preset verslow -tune zerolatency -x265 -params ”crf=QP:keyint=GOP:verbose=1” output.mkv }

\item HM-16.20 (LDP)

\texttt{TAppEncoder -c encoder\_lowdelay\_P\_main.cfg -i input.yuv -DecodingRefreshType=2 -f N -q QP -fr FR -wdt W
-hgt H -IntraPeriod=12 -o output.yuv}

\end{itemize}

For the latest coding standards H.266/VVC \cite{DBLP:journals/tcsv/BrossWYLCSO21}, we use the official reference software VTM-13.2 (LDP) \cite{VTM} as:

\begin{itemize}
\item VTM-13.2 (LDP)

\texttt{EncoderApp -c encoder\_lowdelay\_P\_vtm.cfg
-i input.yuv -DecodingRefreshType=2 -f N -q QP -fr FR -wdt W
-hgt H -IntraPeriod=16 -o output.yuv 
}
\end{itemize} 

In these settings, \texttt{W, H, FR, N, GOP} represent width, height, frame rate, encoded frame numbers and the GOP size, respectively. The quantization parameter (QP) is set as \{21, 23, 27, 31\}. Furthermore, we also discuss the influence of different GOP size in subsequent ablation study.

\emph{\textbf{Coding Performance in  PSNR metric}}: Table \uppercase\expandafter{\romannumeral 1} shows the Bjøntegaard Delta Bit-Rate (BDBR) \cite{Bjontegaard} performance for PSNR metric using the anchor as x265 (veryslow). The lower BD-rate value, the more bit cost reduced,  indicating better coding performance. As shown in the Table \uppercase\expandafter{\romannumeral 1}, our proposed method achieves 37.04\% BD-rate saving on average. On one hand, traditional methods HM-16.20 \cite{HM} and VTM-13.2 \cite{VTM} bring 26.89\% and 49.77\% coding gain on average.  On the other hand, learned methods FVC \cite{DBLP:conf/cvpr/HuL021}, DCVC \cite{DBLP:conf/nips/LiLL21}, SPME(FVC*) \cite{DBLP:conf/mm/GaoC00ZZ22}, SPME(DCVC) \cite{DBLP:conf/mm/GaoC00ZZ22}, CANF-VC \cite{DBLP:conf/eccv/HoCCGP22}, DMVC \cite{DBLP:journals/tcsv/LinJZWMG23} and TCVC \cite{DBLP:journals/tip/JinLPPLL23} save 17.76\%, 22.31\%, 13.13\%, 28.95\%, 25.39\%,26.23\% and 30.66\% BD-rate on average, respectively. It is obvious to see that our scheme is is superior to FVC , DCVC , SPME(FVC*) , SPME(DCVC), CANF-VC , DMVC and TCVC on all benchmark datasets, which proves the strong generalization ability of our proposed MASTC-VC.  Moreover, we also beat popular HM-16.20 by average 10.15\% coding gain. More importantly, we outperform the latest VTM-13.2 on UVG dataset. The corresponding results can also be seen from the RD curves in Fig.9. However, our scheme cannot catch up with the VTM-13.2 in general, although we achieve the SOTA coding performance in the learned methods.

\begin{table*}[htbp]
\centering
\caption{BD-rate(\%) saving comparison between existing baseline methods and the anchor is x265(verslow) in ms-ssim metrics. the best performance of learned methods is marked in bold.}
\label{tab:my-table}
\resizebox{\textwidth}{!}{%
\begin{tabular}{ccccccccccc}
\toprule[1.5pt]
\multirow{2}{*}{Datasets} & \multicolumn{2}{c}{Traditional Methods} & \multicolumn{7}{c}{Learned Methods}                                   \\ \cmidrule(lr{0pt}){2-3}\cmidrule(lr{0pt}){4-11}
 &
  \begin{tabular}[c]{@{}c@{}}HM-16.20 \cite{HM}\cite{HM}\\ (LDP, 4refs)\end{tabular} &
  \begin{tabular}[c]{@{}c@{}}VTM-13.2 \cite{VTM}\cite{VTM}\\ (LDP, 8refs)\end{tabular} &
  \begin{tabular}[c]{@{}c@{}}FVC \cite{DBLP:conf/cvpr/HuL021}\\ (CVPR'21)\end{tabular} &
  \begin{tabular}[c]{@{}c@{}}DCVC \cite{DBLP:conf/nips/LiLL21}\\ (NIPS'21)\end{tabular} &
  \begin{tabular}[c]{@{}c@{}}SPME(FVC*) \cite{DBLP:conf/mm/GaoC00ZZ22}\\ (ACMMM'22)\end{tabular} &
  \begin{tabular}[c]{@{}c@{}}SPME(DCVC) \cite{DBLP:conf/mm/GaoC00ZZ22}\\ (ACMMM'22)\end{tabular} &
  \begin{tabular}[c]{@{}c@{}}CANF-VC \cite{DBLP:conf/eccv/HoCCGP22}\\ (ECCV'22)\end{tabular} &
  \begin{tabular}[c]{@{}c@{}}DMVC \cite{DBLP:journals/tcsv/LinJZWMG23}\\ (TCSVT'23)\end{tabular} &
  \begin{tabular}[c]{@{}c@{}}TCVC \cite{DBLP:journals/tip/JinLPPLL23}\\ (TIP'23)\end{tabular} &
  \textbf{Ours} \\  \midrule
HEVC ClassB               & -14.21             & -41.24             & -47.42 & -44.08 & -40.44 & -47.60 & -48.02 & -51.12 &-58.03 & \textbf{-76.06} \\  \midrule
HEVC ClassC               & -7.93              & -34.18             & -39.11 & -36.30 & -34.49 & -39.00 & -44.76 & -43.52 &-45.60 & \textbf{-55.86} \\  \midrule
HEVC ClassD               & -5.86              & -31.82             & -46.22 & -45.46 & -44.34 & -48.75 & -50.64 & -54.57 &-51.56 & \textbf{-58.43} \\  \midrule
HEVC ClassE               & -27.87             & -57.98             & -59.80 & -35.50 & -34.35 & -44.64 & -44.65 & -58.93 &-54.68 & \textbf{-72.99} \\ \midrule
UVG                       & -14.10             & -41.80             & -42.46 & -42.58 & -32.91 & -49.26 & -48.81 & -49.27 &-48.28 & \textbf{-58.19} \\  \midrule
MCL-JCV                   & -2.40              & -32.36             & -44.54 & -41.38 & -32.22 & -45.03 & -42.86 & -45.56 &-46.98 & \textbf{-61.45} \\  \midrule
AVG                       & -12.06             & -39.90             & -46.59 & -40.88 & -36.46 & -45.71 & -46.62 & -50.50 &-50.86 & \textbf{-63.83} \\ 
\bottomrule[1.5pt]
\end{tabular}%
}
\end{table*}

\begin{figure*}[htbp]
	\centering
	\begin{minipage}{0.32\linewidth}
		\centering
		\includegraphics[width=\linewidth]{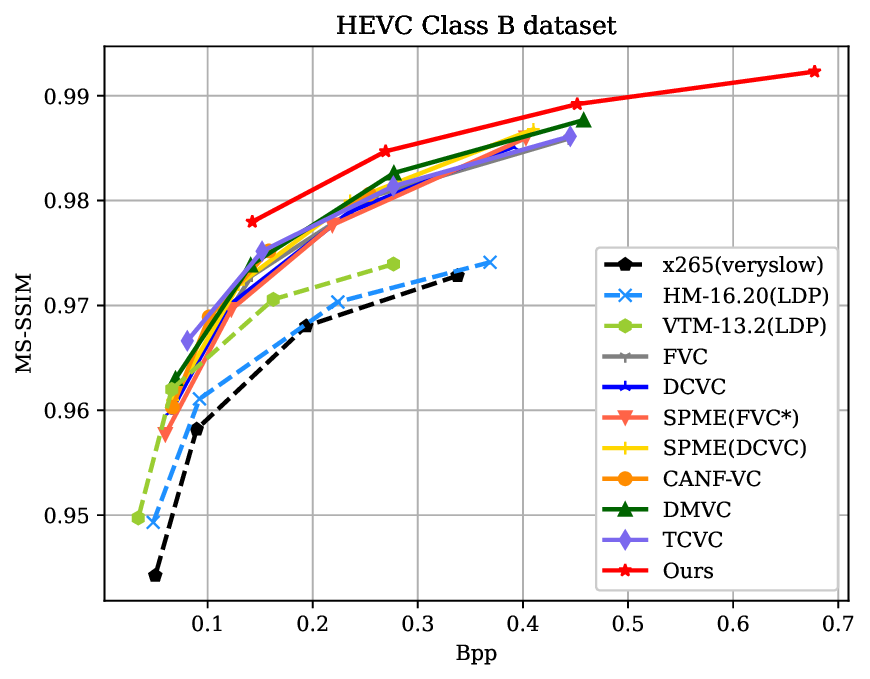}
		\label{chutian1}
	\end{minipage}
	\begin{minipage}{0.32\linewidth}
		\centering
		\includegraphics[width=\linewidth]{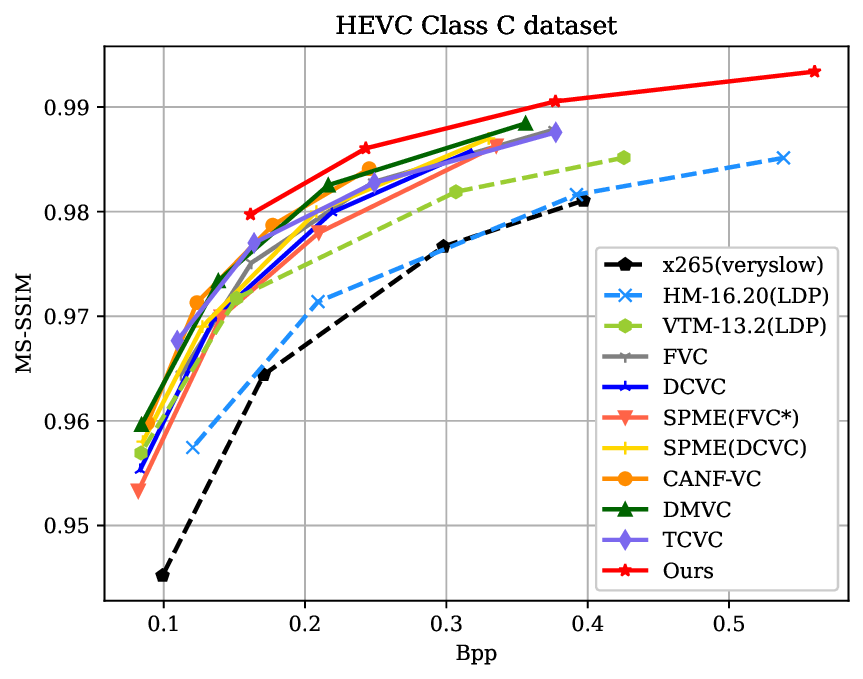}
		\label{chutian2}
	\end{minipage}
	\begin{minipage}{0.32\linewidth}
		\centering
		\includegraphics[width=\linewidth]{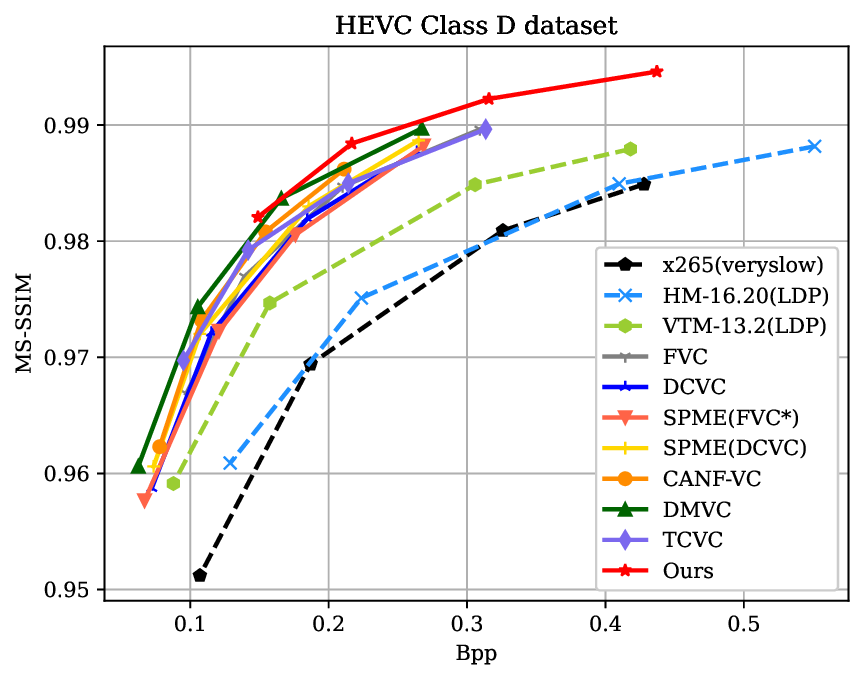}
		\label{chutian2}
	\end{minipage}
	
	\begin{minipage}{0.32\linewidth}
		\centering
		\includegraphics[width=\linewidth]{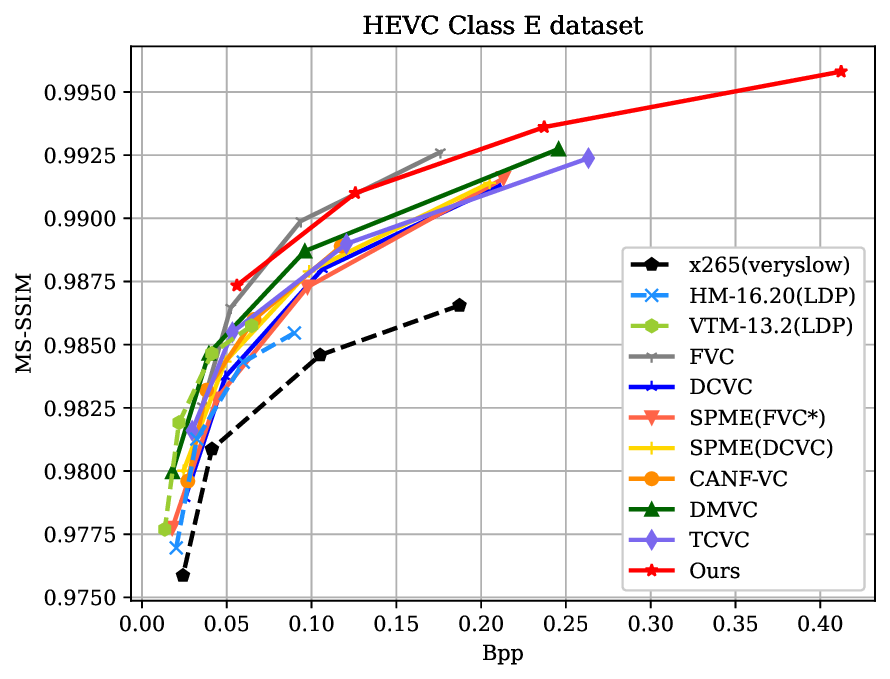}
		\label{chutian3}
	\end{minipage}
	\begin{minipage}{0.32\linewidth}
		\centering
		\includegraphics[width=\linewidth]{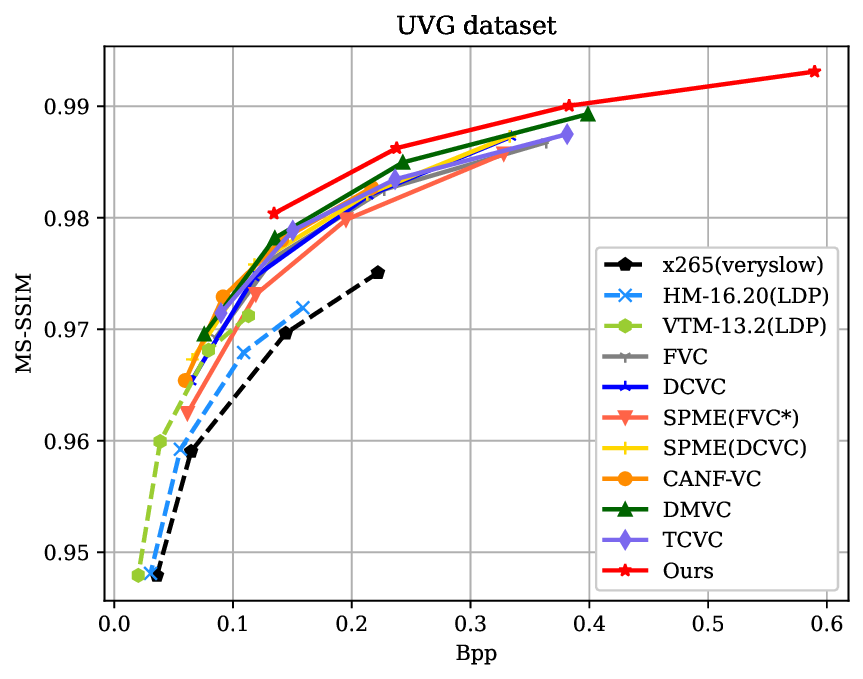}
		\label{chutian4}
	\end{minipage}
    \begin{minipage}{0.32\linewidth}
		\centering
		\includegraphics[width=\linewidth]{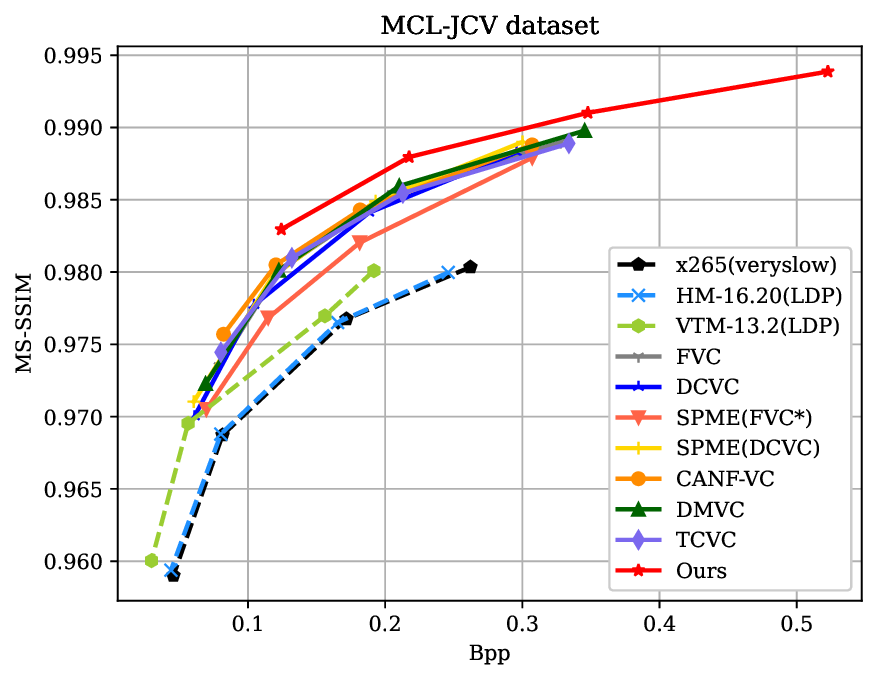}
		\label{chutian4}
	\end{minipage}
 
\caption{Rate-distortion performance evaluation of our proposed method on the HEVC, UVG and MCL-JCV datasets in MS-SSIM metrics.}
\end{figure*}

\emph{\textbf{Coding Performance in MS-SSIM metric}}: Table \uppercase\expandafter{\romannumeral 2} shows the BDBR performance for MS-SSIM metric using the anchor as x265 (veryslow). From Table\uppercase\expandafter{\romannumeral 2}, our method achieves 63.83\% BD-rate reduction on average. Meanwhile, the traditional methods (HM-16.20, VTM-13.2) bring 12.06\% and 39.9\% BD-rate increase on average, respectibely. Also, the learned methods (FVC \cite{DBLP:conf/cvpr/HuL021}, DCVC \cite{DBLP:conf/nips/LiLL21}, SPME(FVC*) \cite{DBLP:conf/mm/GaoC00ZZ22}, SPME(DCVC) \cite{DBLP:conf/mm/GaoC00ZZ22}, CANF-VC \cite{DBLP:conf/eccv/HoCCGP22}, DMVC \cite{DBLP:journals/tcsv/LinJZWMG23} and TCVC \cite{DBLP:journals/tip/JinLPPLL23})  get 46.59\%, 40.88\%, 36.46\%, 45.71\%, 46.62\%, 50.50\% and 50.86\% BD-rate saving on average, respectively. It can be found that the rate-distortion performance of the learned methods is better than the traditional methods, which indicates the potential of data-driven video coding methods in MS-SSIM metric. Furthermore, Our proposed MASTC-VC surpasses the listed learned method \cite{DBLP:conf/cvpr/HuL021, DBLP:conf/nips/LiLL21, DBLP:conf/mm/GaoC00ZZ22, DBLP:conf/eccv/HoCCGP22, DBLP:journals/tcsv/LinJZWMG23,DBLP:journals/tip/JinLPPLL23 }  and traditional methods (HM-16.20 \cite{HM} , VTM-13.2 \cite{VTM} )  by a larger margin on all test datasets, especially on high-resolution (1080p) video datasets. It is important to highlight that our method outperforms the VTM-13.2 by 23.93\% in MS-SSIM metrics. The corresponding results can also be seen from the RD curves in Fig.10.

\begin{figure*}[htbp]
\centering
\includegraphics[width=\linewidth]{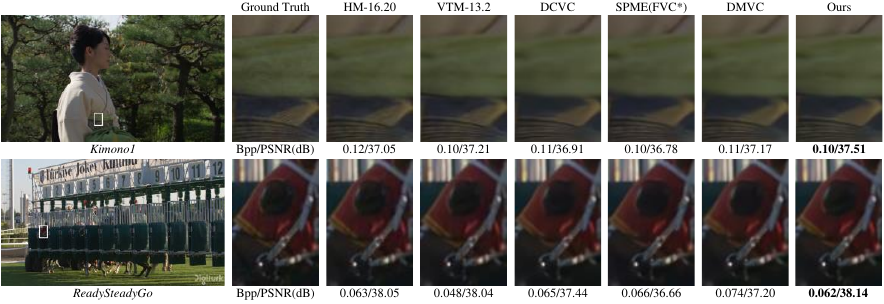}%
\caption{The reconstructed frame from HM-16.20, VTM-13.2, DCVC, SPME(FVC*), DMVC and our proposed method. }
\end{figure*}

\emph{\textbf{Visual Comparison}}:  To further evaluate  the strength of our proposed method (MASTC-VC),  we also provide the visualization of HM-16.20 \cite{HM} , VTM-13.2 \cite{VTM} , DCVC \cite{DBLP:conf/nips/LiLL21}, SPME(FVC*) \cite{DBLP:conf/mm/GaoC00ZZ22}, DMVC \cite{DBLP:journals/tcsv/LinJZWMG23} and our MASTC-VC in Fig.11. It can be obviously seen that our MASTC-VC model can obtain better reconstructed frame quality with less bit consumption than other methods. Specifically, the QP for HM-16.20 and VTM-13.2 is set as 21. We set the $\lambda$ as 2048 for the learned video compression. It is worth noting that our method can retain high-fidelity texture.

\begin{table}[htbp]
\centering
\caption{effectiveness of the proposed different modules of our scheme.}
\label{tab:my-table}
\scalebox{0.9}{
{%
\begin{tabular}{c|c|c|c|c|c|c|c}
\toprule[1.5pt]
MS-MAM & STCCM & B   & C   & D   & E   & UVG & MCL-JCV \\ \hline
\CheckmarkBold      & \CheckmarkBold     & 0.0 & 0.0 & 0.0 & 0.0 & 0.0 & 0.0     \\ \hline
\CheckmarkBold      & \XSolidBrush     & 2.9 & 8.2 & 5.5 & 1.1 & 6.0 & 2.6     \\ \hline
\XSolidBrush & \CheckmarkBold  & 3.7 & 2.6 & 3.3 & 9.1 & 9.7  & 3.3               \\ \hline
\XSolidBrush & \XSolidBrush &  8.3           &  18.9           & 14.0           &   11.6          &  14.2       & 6.6   \\ \bottomrule[1.5pt]
\end{tabular}%
}}
\end{table}

\subsection{Ablation Study}
\emph{\textbf{Effectiveness of the Proposed MS-MAM and STCCM}}: In this work, we pay more attention on motion prediction and contextual coding modules. Hence,  we propose the multiscale motion-aware module (MS-MAM) and spatial-temporal-channel contextual module (STCCM). To validate the contributions of proposed MS-MAM and STCCM, we perform an ablation study as shown in Table \uppercase\expandafter{\romannumeral3}, where the baseline is our full  model (MS-MAM + STCCM).  From the Table \uppercase\expandafter{\romannumeral3} ,we can find that the MS-MAM and STCCM both can improve the compression ratio.  As shown in Fig.12, we also take the \emph{RaceHorses\_416$\times$240\_30} sequence as an example to visualize  inter-frame motion information, reconstructed error  and reconstructed frame quality.  It can be  observed that our proposed MASTC-VC  and "w/ MS-MAM w/o STCCM" model can generate  additional spatial hierarchy in Fig.12 (a), which illustrates the effectiveness of MS-MAM.  Moreover, the proposed MASTC-VC and "w/o MS-MAM w/ STCCM" model reconstruct higher quality frames with less bit consumption in Fig.12 (b) and (c), indicating that the effectiveness of STCCM.

\begin{table}[htbp]
\centering
\caption{Influence of  three different contexts of our STCCM.}
\label{tab:my-table}
\scalebox{0.9}{
\begin{tabular}{l|c|c|c|c|c|c}
\toprule[1.5pt]
                  & B & C & D & E & UVG & MCL-JCV \\ \hline
w/o spatial context &3.6   &3.5   & 2.2  &1.9   & 10.2    & 6.0        \\ \hline
w/o channel context &5.7   & 4.9  &3.7   &2.3   & 11.5    &   9.0      \\ \hline
w/o temporal context &12.3 & 17.2 & 11.5 & 9.8 & 15.4& 9.8 \\ 
\bottomrule[1.5pt]
\end{tabular}}
\end{table}

\emph{\textbf{Influence of Three Different Contexts}}: Furthermore, we aggregate the three contexts to improve the coding performance in spatial-temporal-channel contextual module. We execute ablation studies to explore the influence of three different contexts.  Table \uppercase\expandafter{\romannumeral4} compares the performance influence of spatial, temporal and channel contexts. The baseline is our full  model. It is obvious to see that the three different contexts improve the rate-distortion performance to varying degrees. Among them, the improvement of temporal context is larger than other contexts,  mainly owing to the strong correlation between successive frames.

\begin{table}[htbp]
\centering
\caption{Influence of the channel Grouping of our STCCM }
\label{tab:my-table}
\scalebox{0.9}{
\begin{tabular}{l|c|c|c|c|c|c}
\toprule[1.5pt]
                          & B    & C   & D   & E    & UVG   & MCL-JCV \\ \hline
(16,16,16,16,16,16,16,16) & 3.1  & 1.7 &2.9  & 6.2  & 14.4  &  3.6     \\ \hline
(16,16,16,16,64)          & 3.9  & 2.5 &2.1  & 5.9  & 13.5  &   5.7     \\ \hline
(16,48,64)                & 4.3  &2.3  & 1.7 & 3.3  & 11.6  &  6.8       \\ 
\bottomrule[1.5pt]
\end{tabular}}
\end{table}

\emph{\textbf{Influence of the Channel Grouping}}: To study the influence of the channel grouping of our STCCM, we change the number of channel grouping. The baseline is our final grouping solution. As depicted in Table \uppercase\expandafter{\romannumeral 5}, the BD-rate reduction of our final grouping solution is the highest. Our scheme initially uses fewer channels to assign finer granularity to the beginning chunks and gradually allocates coarser granularity by using more channels, which better fits the latent sapce distribution.

\begin{table}[htbp]
\centering
\caption{The coding performance of different Gop size. The anchor is the GOP10.}
\label{tab:my-table}
\begin{tabular}{l|c|c|c|c}
\toprule[1.5pt]
                  & B & C & D & E  \\ \hline
GOP 4 & 26.6  &23.4 &31.1 & 59.3          \\ \hline
GOP 8&4.3   &2.9  &4.4  & 8.3          \\ \hline
GOP 12& -2.0  &-1.3  & -2.5 &-5.2           \\ \hline
GOP 16 &-3.9  &-2.5  &-4.9  &  -9.7          \\ \hline
GOP 32 &-3.3  & 0.73 &-5.4  & -7.5           \\ 
\bottomrule[1.5pt]
\end{tabular}
\end{table}

\emph{\textbf{Different GOP Size}}: The GOP (group of pictures) is used to alleviate the error propagation. However,  larger GOP sizes will cause reconstructed frame quality decreases and smaller GOP sizes will increase bits cost. We make the experiment to discuss the influence of different GOP size. Table \uppercase\expandafter{\romannumeral 6} shows the RD performance of different GOP size of our proposed method. As we can see from this table, the GOP choice is often a trade-off.  The GOP 16 is the optimal for HEVC Class B, C and E. Yet, GOP 32 is optimal for HEVC  Class D.

\begin{figure*}[htbp]
\centering
\includegraphics[width=\linewidth]{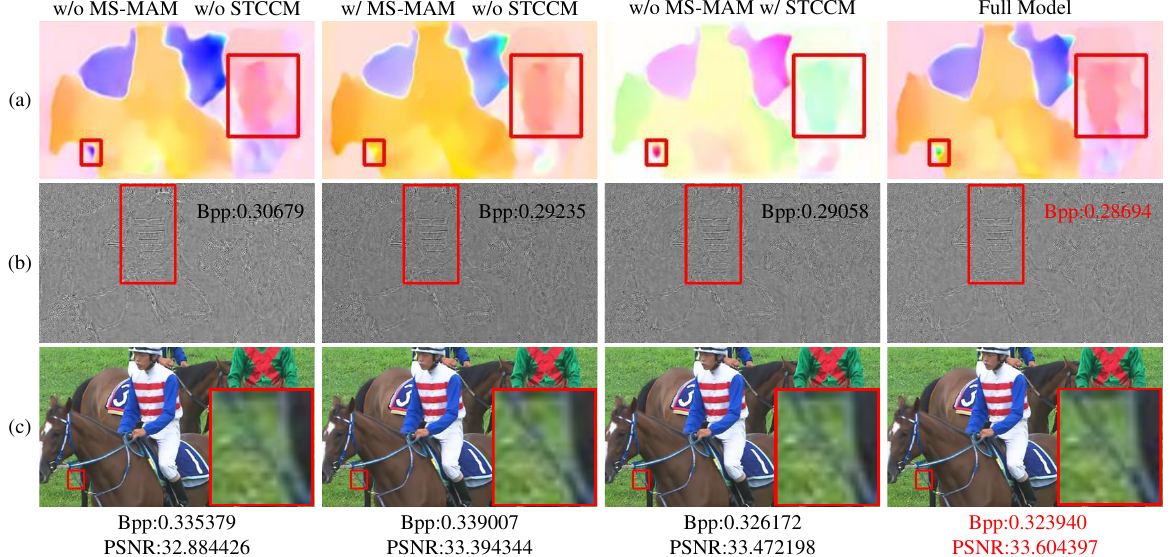}%
\caption{ Visualizing the (a) temporal motion, (b) reconstructed error  and (c) reconstructed frame quality.}
\end{figure*}

\begin{table}[htbp]
\centering
\caption{The rate-distortion performance of different Intra Coding Method in psnr metric.  The anchor is cheng2020\_anchor}
\label{tab:my-table}
\scalebox{0.9}{
\begin{tabular}{c|c|c|c|c|c|c}
\toprule[1.5pt]
                                & B & C & D & E & UVG & MCL-JCV \\ \hline
HM-16.20 intra coding \cite{HM} & 9.7   &-0.1  & 9.1 & 10.5  &17.5      & 19.3        \\ \hline
VTM-13.2 intra coding \cite{VTM} & -2.4  &-8.6  &0.5  &-10.6   & 4.1    &  9.9       \\ \hline
bmshj2018\_hyperprior \cite{DBLP:conf/iclr/BalleMSHJ18} &23.3   &3.0  &11.5  & 37.1  & 18.3    &13.8         \\ \hline
mbt2018 \cite{DBLP:conf/nips/MinnenBT18} &14.8   &-2.5  &4.6  & 14.2  & 10.2    &  6.8       \\ 
\bottomrule[1.5pt]
\end{tabular}}
\end{table}

\subsection{Model Analysis}

\emph{\textbf{Influence of Intra Coding Method}}:  It is well known that I-frames also have an influence on the final encoding performance. Therefore, we also discuss the impact  of different intra-frame coding methods on the overall coding performance. We select the learned image compression methods bmshj2018\_hyperprior \cite{DBLP:conf/iclr/BalleMSHJ18}, mbt\_2018 \cite{DBLP:conf/nips/MinnenBT18},  cheng2020\_anchor \cite{DBLP:conf/cvpr/ChengSTK20}  as well as the traditional methods HM-16.20 intra coding \cite{HM}, VTM-13.2 intra coding \cite{VTM}.  Detailed experimental results are shown in Table \uppercase\expandafter{\romannumeral 7}, where cheng2020\_anchor is used as an anchor. It can be observed  that we can get better RD performance  when more efficient intra-frame coding methods are applied.

\begin{table}[htbp]
\centering
\caption{AVERAGE ENCODING/DECODING Speed (Frames Per Second) and model parameters on 1080p videos}
\label{tab:my-table}
\begin{tabular}{cccc}
\toprule[1.5pt]
Method                                     & Params(M) & Enc Speed & Dec Speed \\ \hline
x265 (veryslow)                            & -         & 0.28      & 19.23          \\
HM-16.20 (LDP) \cite{HM}                   & -         & 0.03      & 10.2        \\
VTM-13.2 (LDP) \cite{VTM}                  & -         & 0.001     & 1.2          \\ \hline
FVC \cite {DBLP:conf/cvpr/HuL021}          & 20.07     & 4.08      & 6.67          \\
DCVC \cite{DBLP:conf/nips/LiLL21}          & 7.94      & 0.082     & 0.028          \\
SPME(FVC*) \cite{DBLP:conf/mm/GaoC00ZZ22}  & 17.75     & 2.57      &  -              \\
CANF-VC \cite{DBLP:conf/eccv/HoCCGP22}      & 31        & 0.625     & 0.95          \\
DMVC  \cite{DBLP:journals/tcsv/LinJZWMG23} & 23.96     & 1.83      & -            \\ 
TCVC \cite{DBLP:journals/tip/JinLPPLL23}   & 28.8      & 0.072     & 0.013          \\ \hline
Ours                                       & 19.87     & 3.24      & 5.32           \\ 
\bottomrule[1.5pt]
\end{tabular}
\end{table}

\emph{\textbf{Computational Complexity}}: We also have measured the computational complexity of our proposed method on NVIDIA RTX 4090 and our model has  19.87M float parameters.  Furthermore, we select the traditional codec (x265, HM-16.20  \cite{HM}, VTM-13.2 \cite{VTM}) and learned methods including FVC \cite{DBLP:conf/cvpr/HuL021}, DCVC \cite{DBLP:conf/nips/LiLL21}, SPME(FVC*) \cite{DBLP:conf/mm/GaoC00ZZ22}, CANF-VC \cite{DBLP:conf/eccv/HoCCGP22}, DMVC \cite{DBLP:journals/tcsv/LinJZWMG23}  and TCVC \cite{DBLP:journals/tip/JinLPPLL23} to encode the 1080p videos. Average encoding and decoding speed is listed in Table \uppercase\expandafter{\romannumeral 8}. 

It is obvious to see that  FVC \cite{DBLP:conf/cvpr/HuL021}  is faster than our method (MASTC-VC), which is owing to the fact that it is based on simple residual coding structures without complex contextual entropy coding networks.  However our proposed method slightly slowers the speed of encoding/decoding and we have achieved higher encoding performance than other learned methods. In the future,  we will design efficient architectures to reduce coding complexity.

\section{CONCLUSION AND DISCUSSION}

In this paper, we propose  a learned video compression (MASTC-VC) via motion-aware and spatial-temporal-channel contextual coding module. In particular, the MS-MAM is proposed to  utilize the multiscale motion prediction information  to estimate spatial-temporal-channel consistent MV in a coarse-to-fine fashion. Furthermore, the STCCM is proposed to  exploit the correlation of latent representation to reduce the bit consumption from spatial, temporal and channel respectively. Lastly, quantitative and qualitative experimental results have demonstrated that our proposed method is superior to previous state-of-the-art (SOTA) methods in terms of both PSNR and MS-SSIM metrics. 


\bibliographystyle{ieeetr}
\bibliography{Ref}

\end{document}